\documentclass[a4paper,10pt]{article}
\usepackage{amsmath,graphicx,color,mathrsfs,subfigure,multicol}

\def\circa#1{\,\raise.3ex\hbox{$#1$\kern-.75em\lower1ex\hbox{$\sim$}}\,}
\makeatletter

\usepackage{epsfig}
\usepackage{cite}
\usepackage{amsmath,amsfonts,amssymb}

\def\be{\begin{equation}}
\def\ee{\end{equation}}

\def\bea{\begin{eqnarray}}
\def\eea{\end{eqnarray}}

\def\beq{\begin{equation}}
\def\eeq{\end{equation}}
\def\baq{\begin{eqnarray}}
\def\eaq{\end{eqnarray}}

\def\hf{\frac{1}{2}}

\def\fnl{f_{\rm NL}}
\def\nnl{n_{\rm f_{NL}}}
\def\fnlpivot{\fnl^{{\rm pivot}}}

\def\fNLlocal{\fnl^{\rm local}}

\def\twop{2\,\mathrm{ perms}}

\def\kpo{\left(\frac{k_1}{k_{\rm p}}\right)}
\def\kpt{\left(\frac{k_2}{k_{\rm p}}\right)}
\def\kpth{\left(\frac{k_3}{k_{\rm p}}\right)}
\def\no{n_{\phi}}
\def\nt{n_{\phi\phi}}

\def\d{{\rm d}}

\def\k{{\vec{k}}}

\def\hf{\frac{1}{2}}

\def\fnl{f_{\rm NL}}
\def\dpo{\delta_1\phi}
\def\dpt{\delta_2\phi}

\begin{document}

\hfill BI-TP 2009/24

\begin{center}
{\LARGE \bf Scale dependence of local $\fnl$}\\[1cm]

{
{\bf Christian T.~Byrnes$^{\,a,\,b\,}$\footnote{{ E-mail}:
byrnes@physik.uni-bielefeld.de}, \,Sami Nurmi$^{\,b\,}$\footnote{{ E-mail}:
s.nurmi@thphys.uni-heidelberg.de
},
 \, Gianmassimo Tasinato$^{\,b\,}$\footnote{{ E-mail}:
g.tasinato@thphys.uni-heidelberg.de
},\, David Wands$^{\,c\,}$\footnote{{ E-mail}: david.wands@port.ac.uk
}
}
}
\\[7mm]
{\it $^a\,$ Fakult{\"a}t f{\"u}r Physik, Universit{\"a}t Bielefeld,
Postfach 100131, 33501 Bielefeld, Germany}

\vskip 0.1cm

{\it $^b\,$ Institut f\"ur Theoretische Physik, Universit{\"a}t Heidelberg,
 Philosophenweg 16 and 19,
69120 Heidelberg, Germany}

\vskip 0.1cm

{\it $^c\,$ Institute of Cosmology $\&$ Gravitation, University of Portsmouth, Dennis Sciama Building, Portsmouth, PO1 3FX, United Kingdom}
\vspace{-0.3cm}

\vspace{1cm}

{\large\bf Abstract}

\end{center}
\begin{quote}

We consider possible scale-dependence of the non-linearity parameter
$\fnl$ in local and quasi-local models of non-Gaussian primordial
density perturbations. In the simplest model where the primordial
perturbations are a quadratic local function of a single Gaussian
field then $\fnl$ is scale-independent by construction. However
scale-dependence can arise due to either a local function of more
than one Gaussian field, or due to non-linear evolution of modes
after horizon-exit during inflation. We show that the scale dependence of  $\fnl$ is typically
first order in slow-roll. For some models this may be observable with experiments
such as Planck provided that $\fnl$
is close to the current observational bounds.


\end{quote}

\bigskip
\bigskip

\noindent

\section{Introduction}

Primordial density perturbations are traditionally described by a Gaussian distribution, characterised by an almost scale-invariant power spectrum. However the detailed information about the primordial density perturbations over a range of cosmological scales offers the opportunity to test in detail the nature of the primordial perturbations, both their scale-dependence and Gaussianity \cite{Komatsu:2008}.

The local model for non-Gaussianity has proved a remarkably popular description of possible deviations from a purely Gaussian distribution of primordial perturbations. In the simplest case the primordial Newtonian potential includes a contribution from both the local value of a linear Gaussian field and a quadratic term proportional to the square of the local value of the Gaussian field:
\be
 \label{originalPhi}
\Phi(\vec{x}) = \varphi_G(\vec{x}) + \fNLlocal \left(
\varphi_G^2(\vec{x}) - \langle\varphi_G^2\rangle \right) \,, \ee
where ${f}^{\rm local}_{\rm NL}$ is a dimensionless parameter
characterising the deviations from
Gaussianity~\cite{Komatsu:2001rj}. $\langle\varphi_G^2\rangle$
denotes the ensemble average, or equivalently the spatial average in
a statistically homogeneous distribution.
Note that following Komatsu and Spergel~\cite{Komatsu:2001rj} we adopt a sign convention for the Newtonian potential $\Phi$ which is the opposite of that used, e.g., by Mukhanov et al~\cite{Mukhanov:1990me}.

In Fourier space we define the power spectrum and bispectrum as
\bea
\langle\Phi_{\vec{k_1}}\Phi_{\vec{k_2}} \rangle = (2\pi)^3 P_{\Phi}(k_1) \delta^{3}(\vec{k_1}+\vec{k_2}) \,,\\
\langle \Phi_{\vec{k_1}} \Phi_{\vec{k_2}} \Phi_{\vec{k_3}} \rangle = (2\pi)^3 B_{\Phi}(k_1,k_2,k_3) \delta^{3}(\vec{k_1}+\vec{k_2}+\vec{k_3}) \,,
\eea
and the amplitude of the bispectrum relative to the power spectrum is conventionally given by the non-linearity parameter
\be
 \label{deffnl}
\fnl(k_1,k_2,k_3)
 \equiv \frac{B_{\Phi}(k_1,k_2,k_3)}{2 \left[ P_{\Phi}(k_1) P_{\Phi}(k_2) + P_{\Phi}(k_2) P_{\Phi}(k_3) + P_{\Phi}(k_3) P_{\Phi}(k_1) \right]} \,.
\ee
In the special case of the local model (\ref{originalPhi}) we have $\fnl=\fNLlocal$ and it is clear that $\fnl$ is, by construction, a constant parameter independent of spatial position or scale.

This local model turns out to be a very good description of non-Gaussianity in some simple physical models for the origin of structure. In particular it can describe the primordial density perturbation on super-Hubble scales predicted, up to second-order, using the $\delta N$-formalism \cite{starob85,SS,LR} if the local integrated expansion is a function of a Gaussian random field, $N(\sigma_g)$, at some initial time, $\sigma_g(\vec{x})=\sigma(t_i,\vec{x})$. We then have
 \be
 \label{Phizeta}
 \Phi = \frac35 \zeta = \frac35 \left[ N - \langle N \rangle \right] \,.
 \ee
A good example is provided by the simplest curvaton scenario~~\cite{Enqvist:2001zp,Lyth:2001nq,Moroi:2001ct}, where quantum fluctuations of a weakly-coupled field during inflation are well described by a Gaussian random field and the primordial density perturbation is determined by the density of the curvaton field when it decays, which is proportional the the square of the initial local field~\cite{Linde:1996gt,Lyth:2002my}.


In this paper we consider extensions of the simple local model (\ref{originalPhi}). In particular we will characterise and quantify the scale-dependence of the parameter $\fnl$ which arise in realistic inflationary models for the origin of structure. Scale-dependence arises due to two key features. Firstly we examine a multi-variate local model where the local expansion is a quadratic function of more than one Gaussian random field. In this case scale-dependent $\fnl$ can arise if the Gaussian fields have differing scale-dependence, leading to a change with scale in the correlation of quadratic terms with the linear perturbation. Secondly, we show that scale-dependent $\fnl$ can arise even when the expansion is a function of a single canonical scalar field.
In this case, it is due to the development of intrinsic non-Gaussianity
associated with the non-linear evolution of the initially Gaussian fluctuations after Hubble-exit.
We will refer to this case as a quasi-local model.

In the case of the so-called equilateral $\fnl$, which arises for
example from DBI inflation (see e.g. \cite{DBI}), the scale dependence has
already been quite well studied both in theoretical models
\cite{Chen:2005fe,Khoury:2008wj,Byrnes:2009qy,Leblond:2008gg} and
 forecasts have
been made for future observational prospects
\cite{LoVerde:2007ri,Sefusatti:2009xu}. For a discussion of the
different possible shape dependences of the bispectrum see
\cite{Fergusson:2008ra}. However for local-type models there has been
little previous consideration of a scale dependence.
The scale dependence of local-type $\fnl$ was first calculated in Byrnes et
al \cite{Byrnes:2008zy} for a specific model of hybrid inflation
with two-fields. It was found that although the scale dependence is
slow-roll suppressed, it depends on a particular combination of
slow-roll parameters which is not negligible and can easily be
larger than the spectral index of the power spectrum. The scale
dependence of $\fnl$ was considered in the case of an ekpyrotic universe in \cite{Khoury:2008wj}. In the case of an exact solution of two-field inflation, which can give rise to a large non-Gaussianity, it was shown that $\fnl$ is scale independent \cite{Byrnes:2009qy}. The observational prospects for local-type models were considered in \cite{Sefusatti:2009xu} which showed that the CMB data is sensitive to a scale dependence of $\fnl$. However they used a very simple Ansatz for the scale-dependent $\fnl$: here we calculate for the first time the full scale dependence of a scale-dependent quasi-local $\fnl$.

We note that higher-order contributions to the primordial
perturbations are expected beyond quadratic order. These only affect
the bispectrum at subleading order in the scalar field
perturbations, although in some cases they might provide the
dominant contribution to observables \cite{Boubekeur:2005fj}. In the
language of \cite{Byrnes:2007tm} they are loop corrections. The
effective scale-dependence of $\fnl$ due to higher-order terms is
examined explicitly in \cite{Kumar:2009ge} and was considered
previously, for example, in \cite{Suyama:2008nt}.  Notice that the
results depend at leading order on the infra-red cut-off, and that
no infra-red complete theory is yet known \cite{Seery:2009hs}. It
would be interesting to further develop  this interesting issue  in
a case in which the infra-red  effects are well-understood.
Note that the loop term is not well described as having constant scale dependence, unlike the cases we consider in this paper. We also note that \cite{Kumar:2009ge} only consider the scale dependence coming
from the logarithmic term which depends on the cut-off, while the other
terms which multiply this will also have a scale dependence in general.
This scale dependence can be calculated using the methods
presented in this paper.

This paper  is organized as follow. In Section \ref{mvarsec},
we focus on multi-variate local models,
which are models that contain contributions to the primordial curvature perturbation from more than one Gaussian field,
in which
 the scale dependence of $\fnl$ is
due to the different scale dependences of the
 fields that drive local expansion. As an example of this we study a mixed inflaton and curvaton scenario in sec.~\ref{sec:inflatonandcurvaton}. In Section \ref{sfsec}, we show
that scale dependence of $\fnl$ can arise also for single field
models, due to the non-linear evolution of initially Gaussian fluctuations
after Hubble exit. In Section \ref{multisec} we extend the discussion
to systems involving multiple fields, taking into
account
the  non-linear evolution of fluctuations after horizon crossing
in this context. We conclude in sec.~\ref{concsec}.

\section{Multi-variate local model}\label{mvarsec}

We will begin by showing how scale dependent $\fnl$ can arise in an idealised model that can describe perturbations from multi-field inflation. A more
systematic discussion on the scale dependence of this quantity, taking into account further second order-effects not considered
here, will be developed in later sections.


A natural extension to the simplest local model given by Eq.~(\ref{originalPhi}) comes from considering a local function of more than one Gaussian field:
\be
 \label{multivariate}
 \Phi(\vec{x}) = \sum_I\varphi^I_G(\vec{x}) + \sum_{I,J} {f}_{IJ} \left( \varphi^I_G(\vec{x}) \varphi^J_G(\vec{x})- \langle\varphi^I_G\varphi^J_G\rangle \right) \,.
\ee
where $\varphi^I(\vec{x})$ describes $n$ independent Gaussian random fields and $f_{IJ}=f_{JI}$ are $n(n+1)/2$ constant parameters.

In Fourier space we have
\be
\langle\varphi^I_{\vec{k}}\varphi^J_{\vec{k'}} \rangle = (2\pi)^3 P_I(k) \delta^{IJ} \delta^{3}(\vec{k}+\vec{k'})
\ee
and all higher-order moments vanish for Gaussian fields.
It is thus straightforward to construct the power spectrum
\be
 P_\Phi(k) = \sum_I P_{\varphi\,I}(k) \,,
 \ee
and the bispectrum
\be
 B_{\Phi}(k_1,k_2,k_3) = 2 \sum_{I,J} f_{IJ} \left[ P_{\varphi\,I}
 (k_1) P_{\varphi\,J}(k_2) + P_{\varphi\,I}(k_2) P_{\varphi\,J}(k_3)
  + P_{\varphi\,I}(k_3) P_{\varphi\,J}(k_1) \right]\,.
 \ee
Thus the non-linearity parameter defined in Eq.~(\ref{deffnl}) is
\be
 \fnl(k_1,k_2,k_3) = \frac{\sum_{I,J} f_{IJ} \left[ P_{\varphi\,I}(k_1) P_{\varphi\,J}(k_2) + P_{\varphi\,I}(k_2) P_{\varphi\,J}(k_3)
 + P_{\varphi\,I}(k_3) P_{\varphi\,J}(k_1) \right]} {\left[ P_{\Phi}(k_1) P_{\Phi}(k_2) + P_{\Phi}(k_2) P_{\Phi}(k_3) + P_{\Phi}(k_3) P_{\Phi}(k_1) \right]} \,.
\ee

If we restrict our attention to equilateral triangles with $k_1=k_2=k_3\equiv k$ we have
\be
 \label{feqmulti}
 \fnl(k) = \sum_{I,J} w_I(k) w_J(k) f_{IJ} \,,
\ee
where the weight given to the non-linearity of each field depends on their contribution to the total power spectrum
\be
 w_I(k) = \frac{P_{\varphi\,I}(k)}{P_{\Phi}(k)} \,.
 \ee
The scale dependence is then given by~\footnote{Notice that
 we follow \cite{Sefusatti:2009xu} but not most previous papers
in defining $\nnl=0$ (as opposed to $-1$) as corresponding to a scale
independent $\fnl$. However we use the notation $\nnl$ rather than
$n_{NG}$ since this can be  easily generalised to consider a scale
dependence of the trispectrum.}
\bea
 \nnl &\equiv& \frac{{\rm d}\ln|\fnl|}{{\rm d}\ln k} \,,\\
  &=& \frac{ \sum_{I,J,K} (\tau_I+\tau_J-2 \tau_K)w_I(k)w_J(k)w_K(k)f_{IJ}} {\sum_{I,J} w_I(k) w_J(k) f_{IJ}} \,,\\
\label{finmultvar}  &=& \frac{ \sum_{I,J} (\tau_I+\tau_J)w_I(k)w_J(k)f_{IJ}} {\fnl} - 2(n-1)\,,
 \eea
where $\tau_I\equiv d\ln P_{\varphi I}/d\ln k$ and the total
spectral tilt is \be n-1 \equiv \frac{{\rm d}\ln P_\Phi}{{\rm d}\ln
k}+3 \,=\,\frac{{\rm d}\ln P_\zeta}{{\rm d}\ln k}+3
 = \sum_I w_I \tau_I+3 \,.
\ee
Although we have here calculated $\nnl$ in the case of an equilateral triangle, we discuss in sections \ref{secgen} and \ref{appB}
  how to treat the more general case.
Note that if all the fields share the same spectral index, $\tau_I=n-4$ for all $I$, then $\nnl=0$. However we will see in Sec.~\ref{multisec} that $\fnl$ could also acquire a scale dependence even if all of the $\tau_I$ are the same,  when
 non-linear evolution of second order fluctuations spoils
the Ansatz (\ref{multivariate}).

\subsection{Mixed inflaton and curvaton scenario}\label{sec:inflatonandcurvaton}

As an example we consider an idealised model of an inflaton, $\phi$,
plus curvaton, $\sigma$, whose large-scale perturbations at some
fixed initial time, after Hubble-exit during inflation, can be
described by Gaussian random fields.
The mixed inflaton-curvaton scenario has previously been studied by several authors~\cite{Lazarides:2004we,Langlois:2004nn,Ichikawa:2008iq,Huang:2008zj,Langlois:2008vk}.
Primordial density perturbations due to adiabatic fluctuations
in the inflaton field during inflation, $\varphi_\phi$, are necessarily very close to Gaussian, $f_{\phi\phi}=0$, and independent of the curvaton perturbations, $f_{\phi\sigma}=0$. However primordial density perturbations due to isocurvature curvaton perturbations may have significant non-Gaussianity~\cite{Lyth:2002my}, $f_{\sigma\sigma}\neq0$. We thus have a bi-variate local model
\be
 \Phi(\vec{x}) = \varphi_\phi(\vec{x}) + \varphi_\sigma(\vec{x}) + f_{\sigma\sigma} \left( \varphi_\sigma^2(\vec{x}) - \langle \varphi_\sigma^2 \rangle \right) \,.
 \ee

The resulting non-linearity parameter for equilateral triangles (\ref{feqmulti}) is
\be
 \fnl(k) = w_\sigma^2(k) f_{\sigma\sigma} \,,
 \ee
where the scale-dependence arises solely due to the scale-dependence
of the weighting function, that  is given by
$w_\sigma(k)=P_\sigma(k)/P_{\Phi}(k)$. We find (here $\tau_\sigma\,=\,
d \ln{P_\sigma}/ d\ln{k}$)
\be
 \nnl = 2(\tau_\sigma+3) - 2(n-1) \,.
 \ee
Using the the results of Wands et al \cite{Wands:2002bn} for the scale dependence of the inflaton and curvaton perturbations, we obtain
\bea
\label{mic:n} n-1&=&
-\left(6-4w_{\sigma}\right)\epsilon+2(1-w_{\sigma})\eta_{\phi\phi}+2w_{\sigma}\eta_{\sigma\sigma}\,, \\
\label{mic:nnl} \nnl &=& 4 \left(1-w_\sigma\right) \left( 2 \epsilon + \eta_{\sigma\sigma}
- \eta_{\phi\phi} \right) \,
 \eea
where, in the notation of \cite{Wands:2002bn},
$\eta_{\sigma\sigma}$, $\eta_{\phi\phi}$ and $\epsilon$ are the
usual slow-roll parameters and $w_\sigma=\cos^2\Delta$ denotes the correlation between the curvaton perturbation, $\varphi_\sigma$, and the primordial curvature perturbation, $\Phi$. Note that in
the limit in which curvaton perturbations dominate over the inflaton in
the primordial density perturbation ($w_\sigma=1$) the
scale-dependence vanishes.

It is interesting to estimate the size of $\nnl$ compared to the
spectral index. Since both are a function of three slow-roll
parameters as well as $w_{\sigma}$, it is not possible to make any
definite relation in general. But in small field models of inflation
one has $\epsilon\ll|\eta_{\phi\phi}|$ and it is reasonable to
assume as well that $\eta_{\sigma\sigma}\ll|\eta_{\phi\phi}|$. Note
that since the quadratic curvaton models requires
$\eta_{\sigma\sigma}>0$ we would otherwise need an unlikely
cancellation between the two $\eta$ parameters in order to have a
red spectral index in agreement with observations. In the case that
$\eta_{\phi\phi}$ provides the dominant contribution to both
(\ref{mic:n}) and (\ref{mic:nnl}) we have independently of
$w_{\sigma}$
\bea\label{nnln} \nnl=-2(n-1)\,. \eea
Observations \cite{Komatsu:2008hk}
tell us that
this implies $\nnl\simeq0.1$ which interestingly is roughly at the
border
of detectability with Planck,
 for a fiducial value $\fnl=50$ \cite{Sefusatti:2009xu}. It should be possible to observationally test the two conditions which are required for (\ref{nnln}) to be valid, by considering additional observables. The tensor-to-scalar ratio is given by $r=16\epsilon(1-w_{\sigma})$ \cite{Wands:2002bn}, and using the $\delta\,N$ formalism it is possible to relate one term of the trispectrum with the bispectrum, as $\tau_{NL}=(6\fnl/5)^2/w_{\sigma}$ (using the notation and formulas of \cite{Byrnes:2006vq}). Hence there are in principle four separate observables which allow one to individually identify the values of the four model parameters in (\ref{mic:n}) and (\ref{mic:nnl}). Finally we note that $g_{NL}$, which parameterises a different term of the trispectrum, does not give an observationally competitive signature in this scenario \cite{Ichikawa:2008iq}.

%

\section{Single field quasi-local models}\label{sfsec}

\label{singlesecf}

We now move towards developing a
systematic formalism to compute the scale-dependence of $f_{\rm
NL}$ in inflationary models. We start by considering the simplest case where the primordial
curvature perturbation is generated from fluctuations in one scalar field. We consider two examples. In the first one, we
discuss a standard single slowly rolling inflaton field with canonical
dynamics. While the non-Gaussianities generated in this model are
unobservably small \cite{Maldacena}, this case serves as a useful
example illustrating the physical origin for the scale dependence of
$\fnl$. As a second example, we consider a curvaton scenario where the effect of inflaton perturbations can be neglected. This can generate observable
non-Gaussianities. We show that while the simplest quadratic curvaton scenario
predicts a scale-independent $f_{\rm NL}$, models with interaction
terms in the curvaton potential typically give rise to
scale-dependence.

We assume the scalar field has the canonical Lagrangian
$$
{\cal L}\,=\,\frac12 \partial_\mu \phi \partial^\mu \phi-V(\phi)\ ,
$$
and that it is light during the inflationary epoch:
 $V''\ll H^2$. We assume
the field obeys slow-roll dynamics during inflation but we do not
require its energy density to dominate the universe.
%
We set $m_{\rm Pl}=(8\pi G)^{-1/2}=1$ and introduce the slow-roll parameters
 \beq
 \label{single_field_sr_parameters}
 \epsilon\,=\,-\frac{\dot{H}}{H^2}\ , \qquad
 \epsilon_{\phi}=\frac12 \left( \frac{V'}{3H^2}\right)^2\ ,\qquad
 \eta_{\phi\phi}=\frac{V''}{3H^2}\ ,\qquad
 \xi^2 \equiv \frac{V'''V'}{9H^4} \ ,
 \eeq
which are all assumed to be small during inflation. If the energy
density of the scalar field $\phi$ dominates the universe during
inflation, we can equate $\epsilon_{\phi}=\epsilon$, but in a more
general set-up these two parameters can differ even at leading order
in slow roll.

We will use the $\delta N$ formalism to characterise the primordial curvature perturbation, given by (\ref{Phizeta}), and analyze the scale dependence of
$\fnl$. The curvature perturbation on superhorizon scales is given by the expansion
  \be
  \label{deltaNzeta}
  \zeta(\vec{x})\,=\,N_{\phi}(t_i)\,\delta \phi(t_i, \vec{x}) +\frac12
  N_{\phi\phi}(t_i)\,\left( \delta \phi(t_i, \vec{x})^2-\langle \delta
  \phi(t_i, \vec{x})^2 \rangle \right)+\dots\ ,
  \ee
where $N(t_i)$ denotes the number of e-foldings from an initial
spatially flat hypersurface at $t_{i}$ to a final uniform energy
density hypersurface, which we assume is chosen such that the
curvature perturbation has frozen to a constant value
$\dot{\zeta}=0$. $N_\phi$ and $N_{\phi\phi}$ denote derivatives with
respect to the initial value of the scalar field $\phi(t_i)$. For
the actual computations, it is convenient to transform
(\ref{deltaNzeta}) into Fourier space writing
  \be
  \label{deltaNzetaft}
  \zeta_{\vec{k}}\,=\,N_{\phi}(t_i)\,\delta \phi_{\vec{k}}(t_i)
  +\frac12 N_{\phi\phi}(t_i)
\left(  \delta \phi \star \delta \phi\right)_{\vec{k}}(t_i)
  +\dots\ ,
  \ee
where $\star$ denotes convolution and subtraction of the zero mode
$\zeta_{\vec{k}=0}$ from (\ref{deltaNzetaft}) is implicitly
understood.

The primordial power spectrum is given to leading order by
 \be
 P_\zeta (k) = N_\phi^2(t_i) P_{\delta\phi}(t_i,k) \,.
 \ee
The calculation of the bispectrum
 and the   derivation of
  the general form for the  $\fnl$
 parameter, defined in eq. (\ref{deffnl}),
 are
 straightforward. One obtains
  \be
 \label{fnlSF}
\frac65 \fnl(k_1,k_2,k_3)\,= \,\frac{ N_{\phi\phi}(t_{\rm i}) \, }{
 N_{\phi}^2(t_{\rm i}) } + \frac{(2\pi)^3\,N_{\phi}^3(t_{\rm i})   \,\,
 B_{\delta \phi}^{c}(k_1, k_2, k_3,t_{\rm i})
  }{{P}_\zeta(k_1){P}_\zeta(k_2)+\twop} \, .
 \ee
The second term on the right is proportional
to the connected part of the scalar field
bispectrum $B_{\delta \phi}^c$, which is
the three-point correlator of the scalar perturbations.

The initial time $t_i$ in (\ref{deltaNzeta}) or~(\ref{deltaNzetaft}) can be
chosen to be any time after the horizon exit of a given mode,
$t_{i}> t_{*}(k)$ where $t_{*}(k)$ is determined by
$k=a(t_*)H(t_*)$. The curvature perturbation is by construction independent of the
choice of the initial time $t_{i}$ \cite{lms}, as discussed in more
detail in Appendix \ref{appA}. However, statistical properties of
the scalar field perturbations $\delta\phi_{\k}(t_i)$ do depend on
the choice of $t_i$.

For a canonical scalar field, slowly rolling during inflation, the
intrinsic non-Gaussianity is slow-roll suppressed at Hubble exit
\cite{Maldacena,SeeryLidsey}. In our quasi-local model we will thus
set the scalar field bispectrum $B_{\delta \phi}^{c}(k_1, k_2,
k_3,t_{\rm i})=0$ when $t_i=t_*(k_i)$ for all $k_i$. However at
later times the distribution in general develops non-Gaussian
features because of the non-linearities of field equations, and it
is this that leads to scale-dependence of $\fnl$.

\subsection{Inflaton field}
\label{sfsr}

The case of single field slow-roll inflation is particularly
instructive, since  there exist general formulas connecting
the derivatives of $N$ with slow-roll parameters.
Moreover, it is possible to appreciate  the connection
between the scale dependence of $\fnl$ and evolution
of the second-order curvature perturbations after horizon exit.

Before focusing  on the properties of $\fnl$, we recall how in the familiar case of a single inflaton field the scale dependence of the power spectrum can be computed in the $\delta N$ formalism in two different ways. Understanding this familiar example will be useful to tackle the more complicated case of the scale dependence of $\fnl$.

The solution of the first order equation of motion for the inflaton perturbations at a
time $t> t_{*}(k)$ soon after horizon exit can be
expressed as \cite{Nakamura}
  \be
  \label{scalsol}
  \delta\phi_{\vec{k}}(t)=\frac{i H(t)}{\sqrt{2 k^3}} \,\left\{
  1+\epsilon+\left[c+\ln{\left(\frac{a(t)
  H(t)}{k}\right)}\right]
 (3\epsilon-\eta_{\phi\phi})
  \right\}a_{\vec{k}}\ ,
  \ee
where $\epsilon=\epsilon_{\phi}$ for the inflaton,  $a_{\vec{k}}$
is a classical random variable satisfying $\langle a_{\vec{k}}
a^\dagger_{\vec{k'}} \rangle =(2\pi)^3\delta^{3}(\vec{k} +\vec{k'})$,
and $c=2-\ln2-\gamma$, with $\gamma$ being the Euler-Mascheroni
constant. To leading order in slow roll, the time dependence of the
slow-roll parameters defined in (\ref{single_field_sr_parameters})
can be neglected.

Using equations (\ref{deltaNzetaft}) and (\ref{scalsol}), one can
work out the dimensionless power spectrum of $\zeta$, defined by
$\langle\zeta_{\vec{k}_1}\zeta_{\vec{k}_2}\rangle=(2\pi)^3\delta(\vec{k}_1+\vec{k}_2)2\pi^2{\cal
P}_{\zeta}(k_1)/k_1^3$. To leading order in perturbations, and next-to-leading order in the slow-roll expansion, this gives
 \beq
 {\cal P}_{\zeta}(k)
 =N_\phi^2(t_i)\frac{H^2(t_i)}{4 \pi^2}\left\{
 1+2\epsilon+2\left[c+\ln{\left(\frac{a(t_i)
  H(t_i)}{k}\right)}\right]
 (3\epsilon-\eta_{\phi\phi})
 \right\}\,,
 \label{parquad}
 \eeq
and the spectral index
  \be
  \label{sidef}
  n-1= \frac{{\rm d}
  \ln {\cal P}_\zeta}{{\rm d} \ln{k}}\ ,
  \ee
can be immediately computed by differentiating (\ref{parquad}) with
respect to ${\rm ln}\, k$. This yields,
  \be
  \label{n_s_single_ti}
  n-1=2\eta_{\phi\phi}-6\epsilon\ .
  \ee

In practice the same result for the spectral index is usually
derived in a different way, that will turn out to be useful for the
following discussion.
In deriving (\ref{parquad}) we have not specified the arbitrary
initial time $t_i$ in (\ref{deltaNzetaft}), except for the technical
constraint that $t_i$ needs to be soon after the horizon exit,
$t_i\gtrsim t_{*}(k)$, for the solution (\ref{scalsol}) to be valid.
A particularly simple choice of $t_i$ is to set it equal to the horizon crossing
for each mode separately $t_i(k)=t_{*}(k)$.

To  obtain the spectral index we can use the identity \cite{SS}
 \be
 \label{imprel}
 H\left( \frac{\partial}{\partial \ln{k}} \right)_
{t_i} = \left(
 \frac{\partial}{\partial t_i } \right)_{k/a}-
 \left(
 \frac{\partial}{\partial t_i}  \right)_{k}
 \ee
applied to eq. (\ref{sidef}),  choosing $t_i$ at the epoch of Hubble
crossing. The second term in the right hand side of (\ref{imprel})
gives zero when applied to the power spectrum, due to the latter
being independent of $t_i$ at fixed $k$. Therefore, at leading order
in the slow-roll expansion, we have
  \bea
  n-1&=& 2\frac{{\rm d}
  \ln{|N_\phi|}}{H\, {\rm d} t_i } + 2\frac{{\rm d} \ln{H}}{H\, {\rm d} t_i }
  \nonumber\\
  &=&
  2\,n_{\phi}
  -2\epsilon\ ,
  \label{fsi}
  \eea
 after defining  $n_{\phi}$
 (and $n_{\phi\phi}$ that will be useful
 in the following) as
 \bea\label{defnphi}
 n_{\phi }&=&\frac{{\rm d} \ln{|N_{\phi}|}}{H\,{\rm d} t_i} \ ,
 \\ \label{defnphiphi}
 n_{\phi\phi}&=&\frac{{\rm d} \ln{|N_{\phi\phi}|}}{H\, {\rm d} t_i} \ .
 \eea
For the inflaton we have $N_\phi=-H/\dot\phi$ and hence
 \bea
\label{defnphisl}
 n_{\phi }&=&
 \eta- 2\epsilon
 \ .
 \eea
Thus the expression (\ref{fsi}) reduces to the usual formula
(\ref{n_s_single_ti}) when focusing on single field slow-roll inflation.

Similarly, we can calculate $\fnl$ in (\ref{fnlSF}) and its scale-dependence using the fact that the curvature perturbation is independent
of $t_i$, see  Appendix \ref{appA}, provided that $t_i$ is chosen after each of the scales $k_j$, $j=1, 2, 3$ has crossed its Hubble scale.
We concentrate here on equilateral triangles, $k_i = k$ postponing
the detailed discussion of non-equilateral configurations until
Appendix~\ref{app:non_eq} (but see also Section \ref{secgen}).

If the field perturbations $\delta\phi_{\bf k}$ are Gaussian at
horizon crossing, that is the bispectrum $B^c_{\delta
\phi}(k,k,k,t_{*})$ vanishes for any $k =a(t_*)H(t_*)$, then the
second term in (\ref{fnlSF}) vanishes at $t_i=t_{*}$. With this
choice, all the scale dependence is encoded in the first term of
(\ref{fnlSF}), which acquires an implicit dependence on $k$ through
the function $t_i = t_{*}(k)$. Note that for the inflaton we have
${\rm d}/H{\rm d}t_i=(\dot\phi/H){\rm d}/{\rm d}\phi$ and hence
 \be
 n_\phi\, =\, - \frac{N_{\phi\phi}}{N_{\phi}^2} \ .
 \ee
Thus, using (\ref{defnphisl}) we have
 \beq\label{fnl1f} \frac65\fnl
 \,=\,\frac{N_{\phi\phi}}{N_{\phi}^2}\, =\, 2\epsilon-\eta\,.
 \eeq
and the scale dependence is given by
 \be\label{connnl}
 \nnl\,=\,\frac{{\rm d} \ln{\left|
 N_{\phi\phi}/N_{\phi}^2
 \right|}
 }{ H\, {\rm d} t_i}
 = n_{\phi\phi}-2 n_{\phi}\,.
 \ee
To leading order in slow-roll, the result can be written explicitly
as
  \be\label{nnlsfi} \nnl\,=\, \frac{
  6\epsilon\eta-8\epsilon^2-\xi^2}{\eta- 2\epsilon} \ .
  \ee
This shows that for single field inflation $\nnl$ is fully
determined by the slow-roll parameters evaluated at horizon crossing
and is first order in slow-roll.
(Note that for $\eta-2\epsilon=0$ the logarithmic scale-dependence diverges simply because $\fnl$ is zero.)

We can physically understand the result by analyzing the evolution
of second order perturbations in this set-up, also using the results
of Appendix \ref{appA}. For single field inflation, we can interpret
the scale dependence of $\fnl$ as being due to the evolution of
second-order perturbations right after horizon exit. Indeed, by
means of the $\delta N$ formalism, splitting scalar perturbations
into first and second order parts as $\delta \phi = \delta_{1}
\phi+\frac12 \delta_{2}\phi$, we can expand the curvature
perturbation up to second order in perturbations as
  \be
  \zeta_{\vec{k}} \,=\,N_{\phi}\,\delta_{1} \phi_{\vec{k}} +\frac12
  \frac{N_{\phi\phi}}{N_\phi^2}\,\left(N_\phi \delta_{1} \phi_{\vec{k}}
  \right)^2+\frac12 N_{\phi}\,\delta_{2} \phi_{\vec{k}}
  \ee
The first two terms contain only the Gaussian first order
perturbation $\delta_{1} \phi $.  The last term, $\frac12
N_{\phi}\,\delta_{2} \phi$, is the second order contribution to the
curvature perturbation that, evolving after the epoch of horizon
crossing $t_*$, is responsible for the scale dependence of local
$\fnl$ in the single inflaton case. In Appendix \ref{appA} we
determine the behavior of $\delta_{2}\phi$, expanding at leading
order in slow roll and around $t_{*}$. Assuming that
$\delta_{2}\phi$ vanishes at $t_{*}$, in accordance with
$\delta\phi$ being Gaussian at this time, we find
 \be
 \label{delta2_sec31}
 \delta_{2} \phi(t)\,=\,\frac{H\,(t-t_{*})}{\sqrt{2\epsilon}}\,\left(8\epsilon^2-6\epsilon\eta+\xi^2\right)
 \left(\delta_{1} \phi_{*}\right)^2
 \,+\,\ldots\ .
 \ee
We see that if $\delta_2 \phi$ does not evolve at the time of
horizon crossing, then $\nnl$ in eq. (\ref{nnlsfi}) vanishes as
expected.

Indeed, instead of setting $t_{i}=t_{*}(k)$ in (\ref{fnlSF}) we
could have derived the result (\ref{nnlsfi}) by treating $t_i$ as
some arbitrary time soon after the horizon exit $t_i>t_{*}(k)$. Even
though we have assumed the perturbations are Gaussian at
horizon-crossing, equation (\ref{delta2_sec31}) explicitly shows
that non-linearities in the field evolution result in a
non-vanishing value for $B_{\delta\phi}^{c}$ at later times, $t_i >
t_{*}$. Since the second term in (\ref{fnlSF}), proportional to
$B_{\delta\phi}^{c}$, depends explicitly on the scale $k$, this
makes $\fnl$ scale-dependent. For singe field inflation, this is
also the only source of scale dependence since the first term in
(\ref{fnlSF}) is independent on time and, with the choice
$t_i>t_{*}(k)$, does not depend on the scale. While this discussion
provides a transparent physical explanation for the origin of the
scale-dependence, in actual computations it is convenient to set
$t_{i}=t_{*}(k)$. These two choices are perfectly equivalent, as
discussed in  Appendix \ref{appB} in more detail.

It is important to emphasize that in this subsection devoted to
adiabatic perturbations of a slowly-rolling inflaton field
we have neglected non-Gaussianity due to the non-linearity of the
inflaton field at horizon-exit which is of the same order as that
which arises due to non-linear evolution after horizon-exit. In any
case $\fnl$ in this model is suppressed by slow-roll parameters and
thus we now turn our attention to models where non-adiabatic field
perturbations can generate much larger non-Gaussianity after
horizon-exit \cite{Langlois:2008vk}.

\subsection{Curvaton scenario}\label{sec:curvaton}

In the curvaton scenario, primordial perturbations are generated by
a curvaton field $\sigma$ which is light and subdominant during
inflation,  and acquires a nearly scale-invariant spectrum of
perturbations. After the end of inflation, the curvaton energy
density dilutes slower than the dominant radiation component. Its
contribution to the total energy density therefore increases and its
perturbations start to affect the expansion history, which generates
curvature perturbations. The curvature perturbations evolve in time,
$\dot{\zeta}\neq0$, until the curvaton eventually decays and the
decay products thermalise with the rest of the universe.  If the
curvaton is still subdominant at the time of decay, the
perturbations can be highly non-Gaussian. Here we concentrate on
this limit and assume that the universe remains effectively
radiation dominated until the curvaton decay, $a\propto t^{1/2}$. In
addition, we assume the primordial perturbation is entirely
generated by the curvaton fluctuations. In this limit the curvaton scenario
effectively reduces to a single field model and it can be analysed using
the formalism developed above.

We consider a generic curvaton potential containing a quadratic term
plus self-interactions
\cite{Dimopoulos:2003ss,Enqvist:2005pg,Enqvist:2008gk,Huang:2008bg,Huang:2008zj,Enqvist:2009zf},
  \beq
  V(\sigma)=\frac{1}{2}m^2\sigma^2+\lambda\frac{\sigma^{n+4}}{M_{\rm P}^n}\
  ,
  \eeq
where $\lambda$ is some coupling and $n\geq0$ is an even integer to
keep the potential bounded from below. We assume the curvaton is
long-lived enough so that the energy density stored in the
interaction term eventually dilutes away and the curvaton ends up
oscillating in an effectively quadratic potential before decaying.
The curvaton energy density at this stage is approximatively given
by
  \beq
  \label{curvaton_rho}
  \rho_{\sigma}\simeq\frac{1}{2}m^2\bar{\sigma}^2(t,\sigma_{*})\equiv \frac{1}{2}\frac{m^2\sigma_{\rm
  osc}^2(\sigma_{*})}{(mt)^{3/2}}\ ,
  \eeq
where $\bar{\sigma}(t,\sigma_{*})$ is the envelope of oscillations
in the quadratic regime. It is proportional to $t^{-3/4}$ and to the
function $\sigma_{\rm osc}(\sigma_{*})$ that depends on the initial
value of the curvaton field $\sigma_{*}$. While $\sigma_{\rm osc}$
is a linear function of $\sigma_{*}$ for a purely quadratic curvaton
potential, it can acquire a highly non-linear form when
self-interactions are switched on. As a representative example, at
the end of this section we will discuss the curvaton model with a
quartic self-interaction term in the potential and the corresponding
form of the function $\sigma_{\rm osc}(\sigma_{*})$.

We assume the curvaton decays instantaneously at $H=\Gamma$, where
$\Gamma$ is the effective curvaton decay rate, and denote its
contribution to the total energy density at this time by
 \beq
  \label{curvaton_r}
  r=\frac{3\rho_{\sigma}}{4\rho_\gamma}\Big|_{\rm dec}\simeq \frac{\sqrt{2}}{4}\,\sigma_{\rm
  osc}^2\left(\frac{m}{\Gamma}\right)^{1/2} \ll 1\ .
  \eeq
Setting $t_i=t_{*}(k)$ in (\ref{deltaNzetaft}) and working to
leading order in $r$, one finds that the derivatives of $N$ in
(\ref{deltaNzetaft}) are given by \cite{LR}
  \beq
  \label{curvaton_Nprimes}
  N_{\sigma}(t_*(k))=\frac{2r}{3}\frac{\sigma_{\rm osc}'}{\sigma_{\rm
  osc}}\ ,\qquad
  N_{\sigma\sigma}(t_*(k))=\frac{2r}{3}\left(\frac{\sigma_{\rm osc}''}{\sigma_{\rm osc}}
  +\left(\frac{\sigma_{\rm osc}'}{\sigma_{\rm
  osc}}\right)^2\right)\ ,
  \eeq
where the primes denote derivatives with respect to $\sigma_{*}$.
Using (\ref{curvaton_Nprimes}), the non-linearity parameter $\fnl$
(\ref{fnlSF}) becomes
  \beq
  \fnl=\frac{5}{6}\frac{N_{\sigma\sigma}}{N_{\sigma}^2}=\frac{5}{4r}\left(1+\frac{\sigma_{\rm osc}\sigma_{\rm
 osc}''}{(\sigma_{\rm osc}')^2}\right)\ .
  \eeq
As before, we have assumed the curvaton perturbations are Gaussian at horizon crossing.

The scale dependence of $\fnl$ can be computed using equation (\ref{connnl})
with $\phi$ replaced by $\sigma$. To find the quantities $n_{\sigma}$ and $n_{\sigma\sigma}$, defined
by (\ref{defnphi}) and (\ref{defnphiphi}), it is convenient to first
express the above results in terms of a reference curvaton value
$\sigma_{1}(\sigma_{*},t_{*}(k))$ measured at some time $t_1$ during
inflation but after the horizon exit of all relevant modes,
$t_1>t_{*}(k)$. Since $\sigma_{\rm osc}(\sigma_{1})$, we can use the
chain rule to write
  \beq
  \label{diff_sigmaosc}
  \frac{\d \sigma_{\rm osc}(\sigma_{1})}{\d t_{*}}
  =\frac{\partial \sigma_{\rm osc}}{\partial\sigma_1}\frac{\d \sigma_{1}}{\d t_{*}}
  =\frac{\partial \sigma_{\rm osc}}{\partial\sigma_1}
  \left(\dot{\sigma}_{*}\left(\frac{\partial \sigma_{1}}{\partial\sigma_{*}}\right)_{t_{*}}
  +\left(\frac{\partial \sigma_{1}}{\partial
  t_{*}}\right)_{\sigma_{*}}\right)\ .
  \eeq
Moreover, as $t_{1}$ is a time event during inflation, the curvaton
evolution until this time is governed by the slow-roll dynamics
  \beq
  \label{slowroll}
  \int_{\sigma_{*}}^{\sigma_1}\frac{\d\sigma}{V'(\sigma)}=-\int_{t_{*}}^{t_1}\frac{\d t}
  {3H(t)}\ ,
  \eeq
from which we get
  \beq
  \label{evaluate_curvaton_derivatives}
  \left(\frac{\partial
  \sigma_{1}}{\partial\sigma_{*}}\right)_{t_{*}}=\frac{V'(\sigma_{1})}{V'(\sigma_{*})}\
  ,\qquad \left(\frac{\partial
  \sigma_{1}}{\partial t_{*}}\right)_{\sigma_{*}}=\frac{V'(\sigma_{1})}{3
  H(t_{*})}\ .
  \eeq
Substituting these into (\ref{diff_sigmaosc}) and using $3
H\dot{\sigma}_{*}=-V'(\sigma_{*})$, we arrive at the result
  \beq
  \frac{\d \sigma_{\rm osc}}{\d {\rm \log}\, k}
  =H(t_{*})^{-1}\frac{\d \sigma_{\rm osc}}{\d t_{*}}=0\ ,
  \eeq
to leading order precision in slow-roll. This derivation can be
immediately generalized for an arbitrary function of the form
$f(\sigma_{1})$, and we arrive at the useful generic result ${\d
f(\sigma_{1})}/{\d {\rm \log}\, k}=0$.

Applying these results to (\ref{curvaton_Nprimes}), we can easily
compute the parameters $n_{\sigma}$ and $n_{\sigma\sigma}$. According to the
discussion above, $\sigma_{\rm osc}$ and $r$ do not depend on the
wavenumber $k$. Similarly, using the chain rule, we can write
$\sigma_{\rm
osc}'=(\partial\sigma_{1}/\partial\sigma_{*})(\partial\sigma_{\rm
osc}/\partial\sigma_{1})$ where the second term depends only on
$\sigma_{1}$ and is constant under differentiation with respect to
$k$. Therefore, we have
  \beq
  \label{curvaton_nphi}
  n_{\sigma}=\frac{\d\ln \sigma_{\rm osc}'}{\d {\rm \log}\, k}=H(t_{*})^{-1}\frac{\d }{\d
  t_{*}}\ln
  \left(\frac{\partial
  \sigma_{1}}{\partial\sigma_{*}}\right)=-H(t_{*})^{-1}\frac{\d \ln V'(\sigma_{*})}{\d t_{*}}=\eta_{\sigma\sigma}\
  ,
  \eeq
where we have used (\ref{evaluate_curvaton_derivatives}) and the
slow roll equation $3H\dot{\sigma}=-V'$. Similarly, we find
  \beq
  \label{curvaton_nphiphi}
  n_{\sigma\sigma}=2 \eta_{\sigma\sigma}+
  \frac{V'''(\sigma_{*})}{3H_{*}^2}\left(\frac{\sigma_{\rm osc}\sigma_{\rm osc}'}
  {(\sigma'_{\rm osc})^2+\sigma_{\rm osc}\sigma_{\rm
  osc}''}\right)\ .
  \eeq
Substituting these into (\ref{connnl}), we find the scale dependence
of $\fnl$ is given by
  \beq
  \label{curvaton_nnl_generic}
  \nnl=n_{\sigma\sigma}-2n_{\sigma}=\frac{V'''(\sigma_{*})}{3H_{*}^2}\left(\frac{\sigma_{\rm osc}\sigma_{\rm osc}'}
  {(\sigma'_{\rm osc})^2+\sigma_{\rm osc}\sigma_{\rm
  osc}''}\right)\ .
  \eeq

It can be immediately seen that $\nnl=0$ for a purely quadratic
curvaton model. However, in the presence of interactions,
$V'''\neq0$, the non-linearity parameter becomes scale-dependent.
Equation (\ref{curvaton_nnl_generic}) gives a rough estimate
$|\nnl|\sim| \eta_{\sigma\sigma}|$ but the precise prediction
depends on the curvaton potential and the details of the dynamics
encoded in the derivatives $\sigma_{\rm osc}'$ and $\sigma_{\rm
osc}''$.

As a specific example, we estimate $\nnl$ for the curvaton potential
$V=1/2\,m^2\sigma^2+\lambda\sigma^4$. This case can be treated
analytically using the results of \cite{Enqvist:2009zf}. In the
limit where the interaction term dominates initially,
$\lambda\sigma_{*}^2\gtrsim m^2$, the function $\sigma_{\rm
osc}(\sigma_{*})$ in (\ref{curvaton_rho}) can be estimated by
equation $(4.9)$ of \cite{Enqvist:2009zf},
  \beq
  \label{curvaton_sigmaosc}
  \sigma_{\rm osc}(\sigma_{*})\simeq\ \sigma_{*}\frac{1.3\,e^{-0.80 \sqrt{s}}
  }{|\Gamma\left(0.75+i\,0.51 \sqrt{s}\right)|}\equiv \sigma_{*}\, \xi(s)\ ,
  \eeq
where $s=\lambda\sigma^2_{*}/m^2$. As discussed in \cite{Enqvist:2009zf}, this agrees well with numerical results for $s\gtrsim 1$
but becomes inaccurate when approaching the quadratic limit $s=0$.
The qualitative behaviour however remains correct even in this limit and we can therefore use (\ref{curvaton_sigmaosc})
for arbitrary $s$, provided that we understand the results in the limit $s\ll 1$ as order of magnitude estimates only.

Substituting (\ref{curvaton_sigmaosc}) in (\ref{curvaton_nnl_generic}), we find
  \beq
  \label{curvaton_nnl_quartic}
  \nnl=\frac{8\lambda\sigma_{*}^2}{H_{*}^2}\left(\frac{\xi(s)^2+2s\xi(s)\xi'(s)}
  {10s\xi(s)\xi'(s)+4s^2(\xi(s)\xi''(s)+\xi'(s){}^2)+\xi(s)^2}\right)
  \equiv\frac{8\lambda\sigma_{*}^2}{H_{*}^2}f(s)=\frac{24s}{1+12s}\eta_{\sigma\sigma}f(s)\ .
  \eeq
The function $f(s)$ can be evaluated using (\ref{curvaton_sigmaosc}) and
its behaviour is illustrated in Fig. 1.
 \begin{figure}[!h]
 \centering
 \includegraphics[width=6.3cm, height=4.2 cm]{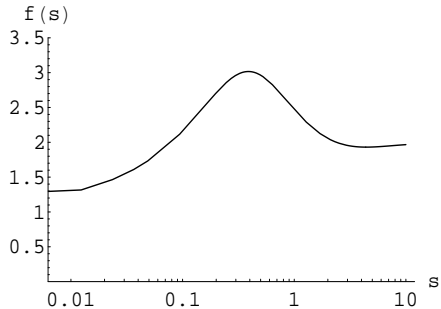}
 \caption{An estimate for the function $f(s)$ in equation
 ({\protect{\ref{curvaton_nnl_quartic}}}) shown as a log-linear plot.
 For $s\lesssim 1$ only a qualitative agreement (error $\sim20\%$) with the exact behaviour is expected.
 }
 \end{figure}
>From (\ref{curvaton_nnl_quartic}) and Fig. 1 we see that $\nnl$ vanishes in the quadratic limit $s=0$ as expected. If
the interaction is initially comparable or dominates over the quadratic term in the potential, $s\gtrsim 1$, we find
$\nnl \sim \eta_{\sigma\sigma}$ up to a factor of few.

Note that for the potential $V=1/2\,m^2\sigma^2+\lambda\sigma^4$
we have $\eta_{\sigma\sigma}>0$ and since
$n-1=-2\epsilon+2\eta_{\sigma\sigma}$, it is assumed in
\cite{Enqvist:2009zf} that $\epsilon$ gives the dominant
contribution to the spectral index. In this case it is clear from
(\ref{curvaton_nnl_quartic}) that $\nnl$ will be subdominant
compared to the spectral index and hence unobservably small. However
in the more general case (\ref{curvaton_nnl_generic}) it is possible
to have $\eta_{\sigma\sigma}<0$ at the time of horizon exit but with
a minimum, e.g.~if the curvaton has an axionic type of potential
\cite{Chingangbam:2009xi}. In this case it might be possible to have
a large $\nnl$. Another case which could lead to a large $\nnl$ in
the curvaton scenario is the non-perturbative curvaton decay
\cite{Enqvist:2008be,Chambers:2009ki} but an investigation of these
points is beyond the scope of this work.

\subsection{Scale-dependence for generic single-field case}\label{secgen}

It is straightforward to generalize the result derived for $\nnl$ to
any quasi-local model where the curvature perturbation is generated
from a single field. Indeed, since equation (\ref{fnlSF}) for $\fnl$
is valid for all such models, we can simply set $t_i=t_{*}$ in it
and read off the result from equations (\ref{defnphi}),
(\ref{defnphiphi}) and (\ref{connnl}) where $\phi$ is now understood
as the generic scalar degree of freedom responsible for generating
the primordial curvature perturbation. We show in Appendix~\ref{appB} that the
result $\nnl=n_{\phi\phi}-2n_{\phi}$ actually holds not only for the
equilateral case but for arbitrary triangles that change their area
while preserving their shape.

In Appendix \ref{appB} we also derive a result that allows to study
variations that change the shape of the triangles. Considering a
triangle with sides $k_1, k_2, k_3$ not too much different from each
other and from a generic reference scale $k_{\rm p}$, we can expand
$\fnl$ around $k_{\rm p}$ as
  \bea
  \label{fnlkdepmt}
   \fnl(k_1,k_2,k_3)=\fnlpivot
  \frac{\kpo^{n-4-\no-2\epsilon_\phi+\eta_{\phi\phi}}\kpt^{n-4-\no-2\epsilon_\phi+
  \eta_{\phi\phi}}\kpth^{\nt+4\epsilon_\phi-2\eta_{\phi\phi}}+\twop}
  {\kpo^{n-4}\kpt^{n-4}+\twop}\ .
  \eea
Here $\fnlpivot$ corresponds to $\fnl$ evaluated on an equilateral
triangle with sides $k_{\rm p}$ and the result is given to leading
order precision in ${\cal
O}(n-1,\epsilon_{\phi},\eta_{\phi\phi},n_{\phi},n_{\phi\phi}) \ln
(k_i/k_{\rm p})$. Note that if $\phi$ is an isocurvature field during
inflation, like in the curvaton scenario
(or more generally in models  characterized
 by a large $\fnl$ \cite{Langlois:2008vk}),
  then $\epsilon_{\phi}\sim
0$ in (\ref{fnlkdepmt}).

The result (\ref{fnlkdepmt}) allows one to generally study the leading
order contributions to the scale dependence of local $\fnl$: $\nnl$
obtained from it reduces to equation (\ref{connnl}) for arbitrary
triangles that change area preserving their shape, while variations
that allow the shape to change in general lead to a different scale
dependence. For example, in the squeezed limit $k_3\ll k_1\simeq
k_2\equiv k$ we find the result
 \bea
 \label{fnlsqueezed}
 \fnl(k,k,k_3)\simeq \fnlpivot
 \left(\frac{k_3}{k_p}\right)^{-\no-2\epsilon_\phi+\eta_{\phi\phi}}
 \left(\frac{k}{k_p}\right)^{\nt-\no+2\epsilon_{\phi}-\eta_{\phi\phi}}\ .
 \eea
If we take a derivative of $\fnl$ with respect to the scale $k$,
while keeping the ratio $k_3/k$ fixed, we again have
$\nnl=\nt-2\no$. But it is clear from (\ref{fnlsqueezed}) that if we
would instead take the derivative w.r.t.~$k_3$ while keeping $k$
fixed or vice versa we would get a different result,
$-\no-2\epsilon_\phi+\eta_{\phi\phi}$ and
$\nt-\no+2\epsilon_{\phi}-\eta_{\phi\phi}$ respectively.

We note that (\ref{fnlkdepmt}) has a more complicated form than the
simple Ansatz for a scale dependent $\fnl$ used by Sefusatti et al
\cite{Sefusatti:2009xu} (Eq.~II.6 of their paper). The two forms
agree in the case of an equilateral triangle but in general it is
not possible to factorise out the local scale dependence which they
denote $F(k_1,k_2,k_3)$. It would be interesting to see how the
forecasted constraints on $\nnl$ vary if one uses the full form for
$\fnl$ given in (\ref{fnlkdepmt}).

\section{Multi field quasi-local models}
\label{multisec}

A set-up in which  more than one scalar field
determines
 the properties of the curvature perturbation
can be treated
 similarly to  single field, but
 in this case there are generically
  two sources which contribute
 to a scale dependence of the quasi-local $\fnl$.
 The first,
 discussed in Section \ref{mvarsec},
 is associated with the fact that the
 various Gaussian
 fields constituting the system can have different
 scale dependence, leading to a change
 with scale in the correlation of quadratic
 terms with the linear perturbations.
 The second, discussed in Section \ref{sfsec},
 is due to the non-linear evolution of second order
 perturbations after horizon exit.

The Lagrangian governing the system reads
\be
{\cal L}\,=\, \frac{1}{2}\,\delta_{IJ}
\, \partial_\mu \phi^I \partial^\mu \phi^J-V(\phi^K)\,.
\ee
Notice that for simplicity we assume a flat metric in field space.  In this
situation,  the expansion for the curvature perturbation,
valid at super-horizon scales,
reads in Fourier space
 \be\label{cruxexp}
 \zeta_{\vec{k}}(t_f)\,=\,N_{I}\delta \phi_{\vec{k}}^{I}(t_i)+\frac12\,N_{IJ}
\left( \delta \phi^{I}\star \delta \phi^{J}\right)_{\vec{k}}(t_i)
 \ee
where we have truncated the expansion at second order.
The
quantities $N_I$, $N_{IJ}$ indicate derivatives of the number of e-foldings
$N=\int_{t_i}^{t_f}\, H dt$ with respect to the scalar fields. In
the framework of  multi-field inflation, they explicitly
depend
on $t_f$, while
 the dependence on $t_i$ is again included in a
 dependence
on $\phi^L$,  writing  $N_{I}\,=\,N_I(t_f, \,\phi^L(t_i))$.
Again,  $\zeta(t_f, \vec{x})$, as defined in (\ref{cruxexp}), does not
depend on $t_i$.

In order to proceed, it is convenient to know some properties of the
solutions for the first order scalar perturbations
soon after horizon exit,
that reads
 \be
 \label{mfscalsol}
 \delta_{\vec{k}} \phi^I(t_i)\,=\,\frac{i
 H(t_i)}{\sqrt{2 k^3}}\,\left\{(1-\epsilon)\delta^I{}_J
 +\left[c+\ln{\left(\frac{a(t_i)
 H(t_i)}{k}\right)}\right]\,\epsilon^I{}_J\right\}\,a_{\vec{k}}^J
 \ee
where $c=2-\ln2-\gamma$ with $\gamma$ the Euler-Mascheroni constant,
while $\epsilon^{IJ}$ is a combination of slow-roll parameters
\cite{Nakamura}
 \be\label{epsilonIJ}
 \epsilon_{IJ}\,=\,\epsilon\,\delta_{IJ}+\frac{V_I
 V_J}{9H^4}-\frac{V_{IJ}}{3H^2}\,,
 \ee
which is small at the epoch of
horizon crossing. Finally, $a_{\vec{k}}^J$ is a classical, time
independent random variable satisfying the orthogonality property $
\langle a_{\vec{k}}^I \,a_{\vec{l}}^J \rangle\,=\,(2\pi)^3\delta^{I
J}\,\delta^3(\vec{k}+\vec{l}) $.

By proceeding as we did in Section  \ref{singlesecf}, we can
calculate the spectral index. By applying  identity
(\ref{imprel}) to the power spectrum, one finds

\be\label{specinm} n-1\,=\,-2\epsilon+  \frac{2\,\sum_I n_I N_I
N_I}{\sum_J N_J N_J} \ee where we introduce the notation (as a
natural extension of (\ref{defnphi}) and (\ref{defnphiphi}))
 \be
 \label{defni}
 n_I\, = \,
 \frac{{\rm d} \ln{N_I}}{H {\rm d}t_i}
 \hskip0.5cm,\hskip0.5cm
 n_{I J}\, = \,
 \frac{{\rm d} \ln{N_{I J}}}{H{\rm d}t_i},
 \ee
and all quantities are evaluated at time $t_i=t_{*}$ corresponding
to horizon exit. We  checked that this expression
coincides with the one of
\cite{SS}, in the case in which the
derivatives of $N$ are explicitly expressed in terms of slow-roll parameters
\cite{VW}.

We can use the above equations and definitions to find an expression
for the  bispectrum
associated with the curvature perturbations,
\bea
B_\zeta&=&(2\pi)^3\,
N_I  N_J  N_L \,\,
B^c_{\delta\phi}(\delta \phi^I_{\vec{k_1}},\,
\delta \phi^J_{\vec{k_2}},\,
\delta \phi^L_{\vec{k_3}})
\, 
\nonumber\\
&+&\,
\frac{
 N_I  N_J  N_{L M} \,\left(\delta^{IL}+q^{IJ}\right)
\left(
\delta^{JM}+q^{JM}\right)}{ N_R N_S N_T N_Z \left(\delta^{TZ}+q^{TZ}\right)
\left(\delta^{RS}+q^{RS}\right)
}\left(\,{P}_\zeta(k_1){P}_\zeta(k_2)
+{\rm perms}\right)\,,
\label{compbis}
\eea
where the scalar perturbations are evaluated at
a common time $t_i$, and $B^c_{\delta\phi}$ is proportional to
the connected part of the three point
function for the scalar perturbations.
The parameter $\fnl$ is obtainable from the definition (\ref{deffnl}), or equivalently
 \be\label{fnl-split} \frac{6}{5} \fnl \,=\,
 \frac{B_\zeta}{\left[{P}_\zeta(k_1){P}_\zeta(k_2)+{\rm
 perms}\right]}.
 \ee

The tensor $q^{IJ}$ is
easily obtained from
  formula (\ref{mfscalsol}).
  For our purposes, it is enough to say that it
  can be expressed as

\be\label{defqij1}
q^{IJ}\,=\,q_{1}^{I J}
+q_{2}^{IJ}\,\ln{\left(\frac{a(t_i) H(t_i)}{k}\right)},
\ee
where $q_1$ and $q_2$ are linear functions of  slow-roll parameters,
with scale-independent coefficients.
  At leading order
in  slow-roll, this quantity depends on $k$
only by means of the logarithm, $\log{k/(aH)}$.

The quasi-local form of $\fnl$ given in eq.~(\ref{fnl-split}) has various
sources of scale dependence. In order to investigate them, we
proceed as in the single-field case. We
 consider
 scale variations that preserve the triangle shape:
here we focus on equilateral triangles, while
the case of
triangles with
different shape is discussed in Appendix {\ref{app:non_eq}}.
In order to calculate $\nnl$, we apply formula (\ref{imprel})
to equation (\ref{fnl-split}). Because $\fnl$ is independent
of $t_i$, the second term of eq (\ref{imprel}) does not
give
contributions. The first term acts only on the first term
of (\ref{fnl-split}), since when applied on the second term it
gives only contributions subdominant in slow-roll
that can be neglected, analogously
to what happens in single field case. Moreover we can neglect its action on $q^{IJ}$, since this provides
contributions that are higher order in slow-roll parameters.

Then, to conclude, the expression for $\nnl$
in this context is

\be\label{nnlmulti}
\nnl\,=\,
\frac{{\rm d}}{H\,{\rm d} \,t_i}\,\ln{ \left[\left( \frac{N^I  N^J
N_{IJ} \,,
}{
\left(N^L N_L\right)^2
}
\right)(t_i)\right]}
\ee
which is a natural generalization of what happens
in the single field case. We make a comparison
with the explicit multi-variate local model discussed in Section \ref{mvarsec}, by expanding
 equation (\ref{nnlmulti})
 by means of the quantities $n_I$
 and $n_{IJ}$ defined in eq. (\ref{defni}).
 One finds
 \bea\label{nnlmultiex}
 \nnl&=&\frac{\sum_{I,J}(n_I+n_J+n_{IJ})N_I N_J N_{IJ}}{\sum_{L,M} N_L N_M
  N_{LM}}-
\frac{4\,\sum_{I}n_{I}N_I N_I}{\sum_JN_J N_J}\ ,\nonumber\\
  &=& \frac{\sum_{I,J}(n_I+n_J+n_{IJ})N_I N_J N_{IJ}}{\sum_{L,M} N_L N_M
  N_{LM}}-2 \left(n_\zeta-1+2\epsilon\right).
\eea
Eq. (\ref{nnlmultiex})  can be seen as the analogue of equation
(\ref{finmultvar}), this time calculated by using the $\delta N$
formalism. If not all of the $n_I$ are the same, there is a
source of scale dependence  for $\fnl$, mainly due to the fact that
first order, Gaussian perturbations  have different
 scale dependence (see Section \ref{mvarsec}).
But,
 even if all the fields
have the same $n_I$, implying $n_{I}\,=\,\frac{1}{2} \,(n_\zeta-1+2 \epsilon)$,
$\nnl$ still will not generically vanish
 due  to the evolution
of  second-order perturbations after horizon
exit (see Section \ref{sfsec}).

\subsection{Multi curvaton scenario}

In this section we consider the possibility that two curvaton fields
contributed to the primordial curvature perturbation. Following
\cite{Assadullahi:2007uw} we assume that the curvatons are
non-interacting and we neglect the inflaton field fluctuations (for
a related scenario see \cite{Huang:2008rj}). In the limit that the
non-Gaussianity is large, which is the observationally interesting
case, we can write the primordial curvature perturbation in the
multivariate local form
\bea\label{mcurvaton:zeta} \zeta=r_a \zeta_a+r_b\zeta_b =
r_a\left(\zeta_{{\rm G}\,a}+\frac34\zeta_{{\rm G}\,a}^2\right) +
r_b\left(\zeta_{{\rm G}\,b}+\frac34\zeta_{{\rm G}\,b}^2\right)\,,
\eea
where $\zeta_a$ is the curvature perturbation of the curvaton field
$a$ and $\zeta_{{\rm G}\,a}$ is the Gaussian part of its
perturbation. We denote the initial, horizon crossing curvaton field
values $a_*$ and $b_*$. Then the Gaussian part of the of the
fluctuation due to the curvaton field $a$ is given by
\bea\label{mcurvaton:zetaGa} \zeta_{{\rm G}\,a}=\frac23\frac{\delta
a_*}{a_*} \eea
and it is useful to further define
\bea \beta=\frac{a_*}{b_*}\,,\qquad r=\frac{r_a}{r_b}\,. \eea
Then
\bea P_{\zeta}&=&r_a^2 P_{\zeta_a}+r_b^2 P_{\zeta_b}=(r_a^2+r_b^2 \beta^2)P_{\zeta_a}, \\
\fnl&=&\frac54\frac{r_a^3+\beta^4r_b^3}{(r_a^2+\beta^2r_b^2)^2}\,. \eea
The full result for $\fnl$ is given in \cite{Assadullahi:2007uw},
but the simplified expression above is a good approximation for
$|\fnl|\gg1$.

Since this form of the curvature perturbation satisfies Ansatz
(\ref{multivariate}) we could have analysed this model using the
results of section \ref{mvarsec}, but we use this model to provide
an example of using the formalism of the previous section. To
calculate the scale dependence we can use (\ref{nnlmultiex}), and
from (\ref{mcurvaton:zeta}) and (\ref{mcurvaton:zetaGa}) we can read
off the derivatives of $N$ as
\bea N_{a}=\frac23\frac{r_a}{a_*}, \qquad N_{aa}=2\frac{r_a}{a_*^2}\,, \eea
and similarly for the derivatives with respect to the $b$ field.
Thus, we can proceed as in Section (\ref{sec:curvaton}) to find
$n_a=\eta_{aa}$, $n_{aa}=2\eta_{aa}$ and similarly for $n_b$.
Putting all of this together we find
\bea\label{mcurvaton:n}
n-1&=&-2\epsilon+\frac{2r^2}{r^2+\beta^2}\eta_{aa}+\frac{2\beta^2}{r^2+\beta^2}\eta_{bb}\,, \\
\label{mcurvaton:nnl}
\nnl &=&4(\eta_{aa}-\eta_{bb})\left[\frac{r^3}{r^3+\beta^4}\frac{\beta^2}{r^2+\beta^2} -\frac{\beta^4}{r^3+\beta^4}\frac{r^2}{r^2+\beta^2}\right].
\eea

Notice that in the limit $\beta^2\rightarrow0$ or
$\beta^2\rightarrow\infty$ only one of the fields contributes to
$\zeta$, since $P_{\zeta_b}=\beta^2 P_{\zeta_a}$. In that case we
can see from (\ref{mcurvaton:nnl}) that $\nnl\rightarrow0$ in
agreement with the expectation from section \ref{sec:curvaton},
because we are considering the non-interacting case. Also in
agreement with Sec.~\ref{mvarsec} we see that $\nnl=0$ if
$\eta_{aa}=\eta_{bb}$. Since both of the terms of $n-1$ which depend
on the $\eta$ parameters must give a positive contribution to the
spectral index and we observe a red spectral index the contribution
from these terms must be small barring  unlikely cancellations
between these two terms and the $-2\epsilon$. In addition each of
the four fractions in (\ref{mcurvaton:nnl}) is less than unity, so
we conclude that (\ref{mcurvaton:nnl}) is likely to be suppressed
compared to the spectral index.


\section{Conclusion}\label{sec:conclusion}\label{concsec}

We have studied generalisations of the local model of
non-Gaussianity, thereby seeing in which cases the standard assumption that
it can be described by a single, scale-independent parameter is valid.
We have shown that this is only strictly valid in specific
models, where the primordial curvature perturbation is generated by
a single scalar field which acts as an isocurvature (``test'') field during inflation
and has a quadratic potential. An example of this is the simplest
curvaton scenario. However as soon as the scalar field has
interactions these generate non-Gaussianity of the field
perturbations after Hubble exit, and these give rise to a scale
dependent non-Gaussianity which we call quasi-local, meaning that in
the limit that the scale dependence of $\fnl$ goes to zero one
recovers the local model.

We have also shown that even in a multi-variate local model where more than one
Gaussian field contributes to the primordial
curvature perturbation the effective $\fnl$ has a scale dependence
unless all fields have the same scale dependence. An example of this is a mixed inflaton and curvaton scenario
or a multi-curvaton model, where the curvatons have quadratic potentials with different masses.


We have developed a formalism, based on the $\delta N$ approach,
which allows us to obtain compact expressions for $\nnl=\,d
\ln{|\fnl|}/d \ln{k}$, in both the single and multi-field case. We
have also discussed more generally the validity of defining a scale
dependence of $\fnl$ with respect to a single scale when the
bispectrum is generally a function of three variables, except in the
case of an equilateral triangle. We have shown that the scale
dependence of $\fnl$ is independent of the shape of the triangle it
describes, provided that we consider scale variations which preserve
its shape. It is then appropriate to perform the simpler
calculation of $\fnl$ for an equilateral triangle, and use this to
calculate $\nnl$ by taking its logarithmic scale dependence.


We applied our formalism to discuss several specific models. Our
results suggest that the scale dependence is typically first order
in slow roll. The precise value however depends quite sensitively on
details of the model which, makes it in principle possible
to use $\nnl$ to discriminate observationally between different models.

We have shown that while the simplest realisation of the curvaton
model has an exactly scale independent $\fnl$, almost any extension
to a more realistic scenario does give rise to a scale dependence.
In the case of an interacting curvaton or a multiple-curvaton
scenario the scale dependence is likely to be suppressed compared to
the spectral index. This is because $\nnl$ in these scenarios
depends schematically on $\eta_{{\rm curvaton}}$ which is normally
positive and, in light of the observational preference for a red
spectrum, this is likely to be small. A detection of both $\fnl$ and
$\nnl$ would therefore be a signal of non-trivial dynamics in the
curvaton scenario, or that the inflaton perturbations have also made
a significant contribution to the observed power spectrum.

In the case of a mixed scenario, in which the primordial curvature
perturbation has contributions from both the inflaton and curvaton
perturbations it is quite natural for there to be a consistency relation
between the scale dependence of the power spectrum and the bispectrum,
$\nnl\simeq-2(n-1)$, and this could be observable with Planck assuming a
large enough fiducial value for $\fnl$, close to the current observational bounds. In this case the bispectrum would
be larger on small scales and this could make a detection of $\fnl$ using
large scale structure data more likely. In any scenario where one includes
the Gaussian perturbations of the inflaton field as well as the
perturbations of a second field which generates non-Gaussianities one will
have a scale dependent $\fnl$, unless both fields have exactly the same
spectral index.

For the case where non-Gaussianity is generated during slow-roll hybrid
inflation one generally has a non-negligble $\nnl$ \cite{Byrnes:2008zy}.
In this case, since the magnitude of $\fnl$ can only grow during inflation, larger
scales, which exit the horizon earlier, will be more non-Gaussian and one
necessarily has $\nnl<0$.

It should be straightforward to extend the formalism we have
presented here to the primordial trispectrum (the four-point
function)
 \cite{Seery:2006js, Byrnes:2006vq}
 and we expect that the two non-linearity parameters which
model it, $\tau_{NL}$ and $g_{NL}$, would also have a first order in
slow roll scale dependence. We have chosen the notation $\nnl$ in a
way that is extendible in an obvious manner to study this. For
example in the case that a single field generates the primordial
curvature perturbation one has a non-Gaussianity consistency
relation, $\tau_{NL}=(6\fnl/5)^2$. Then assuming our formalism can
be extended in the obvious way it follows that
$n_{\tau_{NL}}=2\nnl$.

\section*{Acknowledgements}

It is a pleasure to thank Lotfi Boubekeur, Robert Crittenden, Kazuya
Koyama, Michele Liguori, David Lyth, Sarah Shandera, Emiliano
Sefusatti and Teruaki Suyama for useful discussions and/or comments
on a draft of this work. SN is partly supported by the Academy of
Finland grant 130265. CB is grateful to the University for
Portsmouth and GT is grateful to the University of Bonn for
hospitality during visits when parts of this work were completed.
%

\section*{Appendices}

\appendix

\section{Explicit proof that $\fnl$ is independent of the initial
time}
\label{appA}

In this section we show explicitly that
 $\fnl$ is independent of  $t_i$. In the first part, we focus
 on single   field slow-roll inflation, in which the
 unique scalar plays the role of inflaton field,
  and we study the equations of  motion for
 the perturbations. In the second part, we prove the
 same results using general properties of our definition
 of the curvature perturbation.

In the single field case,
  $\zeta$ is conserved on super horizon scales so we do not need to consider a further dependence on $t_f$.
  Then the $\delta N$ formalism
  provides
  the following expression
\be
    \zeta=\frac{1}{\sqrt{2\epsilon}} \delta\phi +\frac{2\epsilon-\eta}{4\epsilon}\delta^2\phi.
\ee
All quantities on the right hand side are  evaluated at
a given time $t_i$. It is useful to split the scalar field perturbation into first and second order perturbations as $\delta\phi=\dpo+\hf\dpt$. If we choose $t_i$
to coincide with Hubble crossing, $t_i = t_*$, we have
\be\label{fnl1field} \frac65\fnl=\eta-2\epsilon|_{t_*}+\frac65\fnl^{(3)}\,. \ee
We are following the notation of \cite{VW} and $\fnl^{(3)}$ has a complicated $k$ dependence but it is zero if we take
 $\dpt_*=0$
 at horizon
crossing, a condition which we will assume from now on.
 We now  consider what happens if we choose to use a time $t_i>t_*$. We first note that
\be\label{sr-derivative} \frac{d(\eta-2\epsilon)}{dN}=
6\epsilon\eta-
8\epsilon^2-\xi^2\,, \ee
where $\xi^2 \equiv \frac{V'''V'}{V^2}$ is a second order slow-roll parameter. We can get the scalar field time dependence from \cite{Huston}. From (2.13) of that paper it is easy to see that
\be \dpo(t_i)=\dpo_*\left[1+H (t_i-t_\star)(2\epsilon-\eta)\right],
\label{sfsrsol} \ee
We assume that $H (t_i-t_\star)$, corresponding
to the number of e-foldings $N$ between $t_\star$ and $t_i$,
 is much smaller than the inverse of the slow-roll parameters. Since we only need to consider about ten $e$-foldings for observational purposes, this is an adequate approximation.

The evolution of second order perturbations is more complicated, and
we need to consider (2.15) of \cite{Huston},  at leading order in a
slow-roll expansion.
 We find
\be \frac1H\left(\dpt\right)^{{\bf\cdot}}
+(\eta-2\epsilon)\dpt=\frac{V}{V'}\left(8\epsilon^2-6\epsilon\eta+\xi^2\right)\dpo^2. \label{secorel}\ee
We note that the $8\epsilon^2$ term above is not present in (2.19) of \cite{Huston}, which is because they use a non-standard definition of $\epsilon$. For our purposes it is important to include it.
 Since we only need to solve this equation at leading order in slow-roll, we can neglect the
 additional time
  evolution of the slow-roll parameters,
   as well as the evolution of $\dpo^2$. This
  makes the integration simple, and  we find
\be\label{dpt-solution} \dpt(t_i)=\dpt_*+\frac{N}{\sqrt{2\epsilon}}\left(8\epsilon^2-6\epsilon\eta+\xi^2\right)\dpo^2, \ee
For the reason explained above, it does not matter here whether the terms which multiply $N$ are evaluated at $t_i$ or $t_*$.
Using (\ref{fnl-split}), it is then clear that the time dependence of the slow-roll parameters in (\ref{fnl1field}) as given in (\ref{sr-derivative}) is compensated
 by the time dependence of $\dpt$ as given by (\ref{dpt-solution}), rendering $\fnl$ independent of $t_i$.
 The previous discussion can be extended to the case
 of multi-fields.

One can arrive to the same result in a different way, that is
easily
generalizable
the case of multiple scalar fields.
Recall that
the curvature
  perturbation
 $\zeta$
is defined by means
of the spatial part of the metric

\be
g_{ij}\,=\,a^2(t)\,e^{2 \zeta(t,\vec{x})}\,\gamma_{ij}(t,\vec{x})\,,
\ee
where $a(t)$ and $\gamma_{ij}(t,\vec{x})$
indicate background quantities.
 In \cite{lms}, it has been rigorously proved that,
at all orders in the expansion of the perturbations, the curvature
perturbation at superhorizon scales can be expressed in terms
of a variation in the number of e-foldings:
\be\label{zedn}
\zeta\,=\,{\cal N}(t_f, t_i, \vec{x})-N(t_f, t_i)
\,\equiv\,\delta N(t_f, t_i, \vec{x})
\ee
with
\be
N(t_f, t_i)\,=\,\int_{t_i}^{t_f} H dt\,,\hskip0.8cm\hskip0.8cm\,
{\cal N}(t_f, t_i, \vec{x})\,=\,-\frac13 \int_{t_i}^{t_f}\,d t\,
\left(
\frac{\dot{\rho}}{\rho+p}
\right)_{\big| \vec{x}}\,.
\ee
In the previous
equations,
 the energy momentum tensor
  for the system
  is assumed to correspond to an ideal fluid:
  this condition is satisfied for a multiple field
  set-up.

It is simple to see that the curvature
perturbation $\zeta$ given in eq
(\ref{zedn}) is {\it independent} of $t_i$, as long as
$t_i$ labels a flat hypersurface.
Indeed, one has

\be\label{indti}
\frac{\partial}{\partial t_i}\,
\zeta\,=\,-H(t_i)+\frac13\,\left(\frac{\dot{\rho}}{\rho+p}\right)(t_i, \vec{x})\,.
\ee
On the other hand, as discussed in eqs (18) and following
of \cite{lms}, as long as one is focusing on flat hypersurfaces,
the following energy conservation equation holds
\be
\frac{\dot{a}}{a}\,=\,-\frac{\dot{\rho}}{3(\rho+p)}
\ee
Using the previous relation, (\ref{indti}) ensures that $\zeta$ is independent
of $t_i$, at all orders in the expansion, as we wanted to prove.
Because the curvature perturbation is independent
of $t_i$, $\fnl$ is also independent
of this quantity.

\smallskip

The information that $\zeta$ is independent of $t_i$,
allows also to straightforwardly obtain the equations of motion for
first and second order scalar perturbations $\delta_1 \phi$ and
$\delta_2 \phi$.  Expanding $\zeta$ at first order in perturbations,
one has \be \zeta (\vec{x})\,=\,N_\phi(t_i)\,\delta_1
\phi(t_i,\vec{x}) \ee
The fact that the right hand side is
independent of $t_i$ provides the equation (we use the notation of
Section \ref{sfsec}) \be \left(\delta_1
\phi\right)^{\cdot}\,+\,n_\phi\,H\,\delta_1 \phi\,=\,0 \ee leading
to eq. (\ref{sfsrsol}) for single field slow roll inflation.
Analogously, expanding $\zeta$ up to second order and proceeding in
the same way, one finds the following equation of motion for second
order perturbations, valid at leading order in slow roll \be
\left(\delta_2 \phi\right)^{\cdot}+n_\phi\, H\,
\delta_2\phi\,=\,H\,N_{\phi}\,n_\phi\,\left(n_{\phi\phi}-2 n_{\phi}
\right)\,\left(\delta_1\phi\right)^2 \,, \ee that corresponds to
equation (\ref{secorel}). It should be possible to extend the present method to the multi-field case, even with non-canonical kinetic
terms, possibly reproducing the results of \cite{RenauxPetel:2008gi}.

\section{Non-equilateral triangles}
\label{app:non_eq}\label{appB}

In this appendix, we explain how to calculate
the running of $\fnl$ for generic triangle
configurations. We directly  discuss a multi-field
system, the single field
case is easily obtained
as special limit of our computations.
 To perform the calculation, it is convenient
to set the initial time $t_i$,
in the expansion
 (\ref{cruxexp}) for $\zeta_{\bf k}$, equal to
the horizon crossing time $t_{i}=t_{*}(k)$. With
this choice and using Wick's theorem, the bispectrum can be written as
  \be
  \label{bispectrum_multi_vw_1st}
  B_{\zeta}(k_1,k_2,k_3)=
  N_{(1) I}  N_{(2) J}  N_{(3)
   L M}\int \frac{{\rm d}{\bf
  q}}{(2\pi)^3}\langle\delta\phi^{I}_{{\bf k}_1}(t_1)\delta\phi^{L}_{{\bf
  q}}(t_3)\rangle \langle\delta\phi^{J}_{{\bf k}_2}(t_2)\delta\phi^{M}_{{\bf
  k}_3-{\bf q}}(t_3)\rangle+\twop\ ,
  \ee
since at horizon crossing we set the connected three and
four point functions of $\delta\phi$ to zero, thereby assuming the perturbations are
Gaussian at this time. In the previous formula, we introduced the
notation $N_{(j) I}\equiv N_{I}(t_f,t_j)$ with $t_j$ corresponding
to the horizon crossing epoch for the mode $k_j$. Notice that the
two point functions in (\ref{bispectrum_multi_vw_1st}) involve
arguments evaluated at different times. In general, to first order
in slow-roll, from (\ref{mfscalsol}) it follows that
  \be
  \label{delta_phi_evol_appb}
  \delta \phi_{\vec{k}}^L (t_a)\,=\,\left[ \delta^L{}_{M}
  +
  \left(
  \ln{\frac{a(t_a) H(t_a)}{a(t_b) H(t_b)}}\right)\,\hat\epsilon^{L}{}_{M}\right]
  \,  \delta \phi_{\vec{k}}^M (t_b)\ .
  \ee
Here we have defined a quantity
  \beq
  \hat\epsilon_{L M}\,=\,
  \frac{V_L V_M}{9H^4}-\frac{V_{LM}}{3H^2}\,,
  \eeq
which differs by a factor $2\epsilon\delta_{LM}$ from
$\epsilon_{LM}$ defined in (\ref{epsilonIJ}) since we have
explicitly included the time evolution of $H$ in
(\ref{delta_phi_evol_appb}). Using the relation between time and
scale at horizon crossing, $k=aH$, to leading order in slow-roll the
two-point functions with unequal time arguments can be written as
  \baq
   \label{twopoint_multi_vw}
  \langle\delta\phi^{I}_{{\bf k}_1}(t_1)\delta\phi^{L}_{{\bf
  q}}(t_3)\rangle&=&\left[
  \delta^{L}{}_{M}+\left({\rm
  ln}\frac{k_3}{k_1}\right)\,\hat\epsilon^{L}{}_{M}\right]\,
  \langle\delta\phi^{I}_{{\bf k}_1}(t_1)\delta\phi^{M}_{{\bf
  q}}(t_1)\rangle\\\nonumber
  &=&\left[\delta^{L}{}_{M}+\left(
  {\rm
  ln}\frac{k_3}{k_1}\right)\,\hat\epsilon^{L}{}_{M}\right]
  \,(2\pi)^3\delta^{(3)}({\bf k}_1+{\bf
  q})\, \frac{(\delta^{IM}+2c\hat\epsilon^{IM})P_{\zeta}(k_1)}{N_{(1)
   R}N_{(1)
    S}(\delta^{RS}+2c\, \hat\epsilon^{RS})}\\
  \nonumber
  &=&(2\pi)^3\delta({\bf k}_1+{\bf
  q})\frac{\left[\delta^{IL}+\left(2c+{\rm ln}\frac{k_3}{k_1}\right)\hat\epsilon^{IL}\right]P_{\zeta}(k_1)}
  {N_{(1) R}N_{(1) S}(\delta^{RS}+2c\,\hat\epsilon^{RS})}\ ,
  \eaq
and similarly for the other terms in (\ref{bispectrum_multi_vw_1st}).
Recall that $c=2-\ln2-\gamma$, with $\gamma$ the Euler-Mascheroni
constant. The result is independent of the time at which the
slow-roll matrix $\hat\epsilon^{IJ}$ is evaluated since we work to
leading order in slow-roll. Using (\ref{twopoint_multi_vw}), the
bispectrum (\ref{bispectrum_multi_vw_1st}) can be written as
  \baq
  \label{bispectrum_multi_vw}
  B_{\zeta}(k_1,k_2,k_3)&=&\left\{\frac{
  N_{(1)I}  N_{(2) J}  N_{(3)  L M}\left[\delta^{IL}+\left(2c +{\rm ln}\frac{k_3}{k_1}\right)\hat\epsilon^{IL}\right]
  \left[\delta^{JM}+\left(2c +{\rm ln}\frac{k_3}{k_2}\right)\hat\epsilon^{JM}\right]}
  { N_{(1)R} N_{(1)S} N_{(2) T} N_{(2) Z} \left(\delta^{RS}+2c\hat\epsilon^{RS}\right)  \left(\delta^{TZ}+2c\hat\epsilon^{TZ}\right)}\right\}
  \nonumber \\ &&
  \times{P}_\zeta(k_1){P}_\zeta(k_2)
  +\twop\ ,
  \eaq
corresponding to equation (\ref{compbis}) rewritten using the choice
$t_{i}=t_{*}(k_i)$ in (\ref{cruxexp}) for each of the three modes
$k_i$. From (\ref{fnl-split}) one finds the corresponding
non-linearity parameter $\fnl$.

To evaluate the scale dependence of $\fnl$, we perform an expansion
around a pivot scale $k_{\rm p}$ not too different from $ k_i$'s.
Let us start by considering scale variations that preserve the
(arbitrary) shape of the triangles: we then write the wavenumbers
$k_i$ in terms of dimensionless parameters $\alpha_{i}$ and a common
scale $k_i=k\alpha_i$. Here we treat the parameters $\alpha_i$ as
constants, concentrating on variations of the scale $k$ that
preserve the shape of the triangle. Denoting the derivatives of $N$
at the pivot scale by
  \beq
  n_{I}
  =\frac {{\rm d\, ln}\, N_{I}}{H{\rm d}t}\Big|_{a=k_{\rm p}/H}\ ,
  \qquad n_{IJ}
  =\frac {{\rm d\, ln}\,
  N_{IJ}}{H{\rm d}t}\Big|_{a=k_{\rm p}/H}\ ,
  \eeq
and defining
  \beq
  \hat{n}=\frac{\sum_{I}n_{I}N_I N_I}{\sum_JN_J N_J}\ ,
  \eeq
we can expand $f_{\rm NL}$ around  $k_p$ as
  \beq
  \label{fnl_expansion}
  f_{\rm NL}(k)=f_{\rm NL}(k_{\rm
  p})\left[1+\left(\frac{\sum_{I,J}(n_I+n_J+n_{IJ})N_I N_J N_{IJ}}{\sum_{L,M} N_L N_M
  N_{LM}}-4\hat{n}\right){\rm ln}\left(\frac{k}{k_{\rm p}}\right)+\ldots\right]\ .
  \eeq
Here the coefficient of the logarithm is given to leading order in
slow-roll parameters and to the same precision the non-linearity
parameter at the pivot scale reads $f_{\rm NL}(k_{\rm p})=(5/6) N_I
N_J N_{IJ}/(N_L N_L)^2$. {}From equation (\ref{fnl_expansion}) we
see that for a generic non-equilateral configuration with a fixed
shape, the scale dependence of the non-linearity parameter is given
by
  \beq
  \label{n_nl_multi_non_eq}
  \nnl=\frac{\sum_{I,J}(n_I+n_J+n_{IJ})N_I N_J N_{IJ}}{\sum_{L,M} N_L N_M
  N_{LM}}-4\hat{n}\ ,
  \eeq
to leading order precision in slow-roll and in the expansion in
${\rm ln}\,(k/k_{\rm p})$. To this precision,  the result coincides
exactly with the one obtained for the equilateral case, equation
(\ref{nnlmultiex}), and is independent of the constants $\alpha_i$
parameterising the shape of the triangle.

However, if both the scale and the shape of the triangle vary
simultaneously, the result will in general be different. In
principle, it is straightforward to work out the scale dependence
for an arbitrary variation, starting  from the general equation
(\ref{bispectrum_multi_vw}). As an example we discuss the single
field case. The bispectrum (\ref{bispectrum_multi_vw}) depends on
$\hat\epsilon_{I J}$ which in the single field case has only one
component
$\hat\epsilon_{\phi\phi}\,=\,2\epsilon_\phi-\eta_{\phi\phi}$.
Equation (\ref{bispectrum_multi_vw}) can be expanded around the
pivot scale $k_p$ as
 \be
 B(k_1, k_2, k_3)=\frac{N_{\phi\phi}}{N_{\phi}^2}\,
  \left[ 1+ \left(n_{\phi\phi}+2\hat\epsilon_{\phi\phi}\right)
  \ln\left({\frac{k_3}{k_p}}\right)
  -  \left(n_{\phi}+\hat\epsilon_{\phi\phi}\right)
  \ln{\left(\frac{k_1}{k_p}\right)}- \left(
  n_{\phi}
  +\hat\epsilon_{\phi\phi}\right)
  \ln{\left(\frac{k_2}{k_p}\right)}
  \right] \,{P}_\zeta(k_1){P}_\zeta(k_2)\nonumber\ ,
 \ee
plus permutations. To the same order of accuracy as in
(\ref{fnl_expansion}), we can then expand $\fnl$ around the pivot
scale as
  \bea
  \label{fnlkdep}
   \fnl(k_1,k_2,k_3)=\fnlpivot
  \frac{\kpo^{n_{\zeta}-4-\no-2\epsilon_\phi+\eta_{\phi\phi}}
  \kpt^{n_{\zeta}-4-\no-2\epsilon_\phi+\eta_{\phi\phi}}\kpth^{\nt+4\epsilon_\phi-2\eta_{\phi\phi}}+\twop}
  {\kpo^{n_{\zeta}-4}\kpt^{n_{\zeta}-4}+\twop}\ .
  \eea
This expression allows to study the scale dependence of $\fnl$ in
full generality for arbitrary variations of the triangle.

\end{document}